\documentclass[aps,pra,twocolumn,superscriptaddress,nofootinbib]{revtex4-2}

\usepackage[dvipsnames]{xcolor}
\usepackage{hyperref}
\hypersetup{
  breaklinks=true,
  colorlinks=true,
  allcolors=BlueViolet,
}

\usepackage{graphicx} 

\usepackage{orcidlink}  
\usepackage{amsmath}
\usepackage{physics,braket,bm}
\usepackage[normalem]{ulem}

\newcommand \be{\begin{equation}}
\newcommand \ee{\end{equation}}

\newcommand \bea{\begin{eqnarray}}
\newcommand \eea{\end{eqnarray}}

\usepackage{amssymb}

\usepackage[inline]{enumitem}
\setlist[enumerate]{leftmargin=*}

\newcommand{\InfleqtionM}{Infleqtion, Madison, WI, 53703, USA}
\newcommand{\InfleqtionB}{Infleqtion, Boulder, CO, 80301, USA}
\newcommand{\UWM}{Department of Physics, University of Wisconsin-Madison, 1150 University Avenue, Madison, WI, 53706 USA}
\def \addCQuIC {Center for Quantum Information and Control, University of New Mexico, Albuquerque, 87131, NM, USA}
\def \addPandAUNM {Department of Physics and Astronomy, University of New Mexico, Albuquerque, NM, 87106, USA}

\newcommand {\rsub}[1]{\textcolor{black}{#1}}

\newcommand {\rrsub}[1]{\textcolor{black}{#1}} 

\begin{document}

\title{An asymmetric and fast Rydberg gate protocol for \rsub{entanglement outside of the blockade regime}}
\author{Daniel C. Cole}
\affiliation{\InfleqtionB}
\author{Vikas Buchemmavari\orcidlink{0000-0002-1592-5626}}
\affiliation{\addCQuIC}
\affiliation{\addPandAUNM}
\author{Mark Saffman}
\affiliation{\InfleqtionM}
\affiliation{\UWM}
\date{\today}

\begin{abstract}
We analyze a new Rydberg gate design based on the original $\pi-2\pi-\pi$ protocol [Jaksch, et. al. Phys. Rev. Lett. {\bf 85}, 2208 (2000)] that is modified to enable high fidelity operation without requiring a strong Rydberg interaction. 
The gate retains the $\pi-2\pi-\pi$ structure with an additional detuning added to the $2\pi$ pulse on the target qubit. The protocol reaches within a factor of 2.39 (1.68) of the fundamental fidelity limit set by Rydberg lifetime for equal (asymmetric) Rabi frequencies on the control and target qubits. 
We generalize the gate protocol to arbitrary controlled phases.
We design optimal target-qubit phase waveforms to generalize the gate across a range of interaction strengths and we find that, within this family of gates, the constant-phase protocol is time-optimal for a fixed laser Rabi frequency and tunable interaction strength.  \rsub{Quantum control techniques}   are used to design gates that are robust against variations in Rydberg Rabi frequency or interaction strength. 
\end{abstract}

\maketitle

\section{Introduction}

The strong interaction of Rydberg atoms provides a mechanism for entangling neutral atom qubits as was originally proposed in Ref. \cite{Jaksch2000} using a $\pi-2\pi-\pi$ sequence of ground-Rydberg pulses. The original protocol was first demonstrated in Refs. \cite{Isenhower2010,Zhang2010}, but has since been supplanted by improved protocols that apply the same optical pulses symmetrically to both atoms \cite{YSun2020,Saffman2020,Levine2019,Robicheaux2021,Jandura2022, Pagano2022, Mohan2023}.  The symmetric protocols simplify optical control requirements and use of a time-optimal protocol has led to  $\sf CZ$ fidelities surpassing  ${\mathcal F}=0.993$ with four different atomic species: Rb \cite{Evered2023}, Cs \cite{Radnaev2025}, Sr \cite{Tsai2025}, and Yb \cite{Peper2025,Muniz2025,Senoo2026}.
 While high fidelity is critical for quantum error correction, leveraging the long range nature of the Rydberg interaction is also of interest for implementing non-local error correcting codes \cite{Poole2025a,Pecorari2025}, \rsub{as well as gates between logical qubits, without physical reconfiguration of the qubit array which may come with downsides such as increased circuit runtime and atom heating.} \rsub{To avoid these potential drawbacks, architectures with individually addressed Rydberg excitation (alone or in combination with global interaction zones) may be considered \cite{Graham2022, Radnaev2025,Saffman2025}, and in this case gate protocols need not be limited to symmetric Rydberg excitation.}

We present here a modified version of the $\pi-2\pi-\pi$ gate that has the  feature of having no coherent rotation error (that is, it achieves unity fidelity in the absence of dissipation and experimental imperfections), even in the partial blockade limit  of interaction strength $V$ comparable to Rydberg excitation rate $\Omega$. The ability to provide high fidelity for moderate $V$ implies that the gate can be operated at large inter-atomic spacings. As we discuss, the Rabi rate of the control qubit Rydberg coupling need not be limited by the blockade strength, which allows the gate to reach shorter durations than some state-of-the-art implementations.
\rrsub{A recent experimental realization of this specific modified protocol was reported in Ref.~\cite{Cole2023}. The present theoretical investigation establishes the fundamental fidelity bounds and speed-limits of this gate. The fidelity achieved in Ref. \cite{Cole2023} was limited by technical noise sources but establishing these theoretical limits is essential for assessing the gate's viability for future hardware-efficient approaches to neutral atom based  quantum computation. We note that the asymmetric protocol is directly applicable to two-species Rydberg gates \cite{Anand2024,Miles2026}, as has been shown in a recent demonstration \cite{YuWang2026}.}

\begin{figure*}[!t]
\centering
 \includegraphics[width=16.cm]{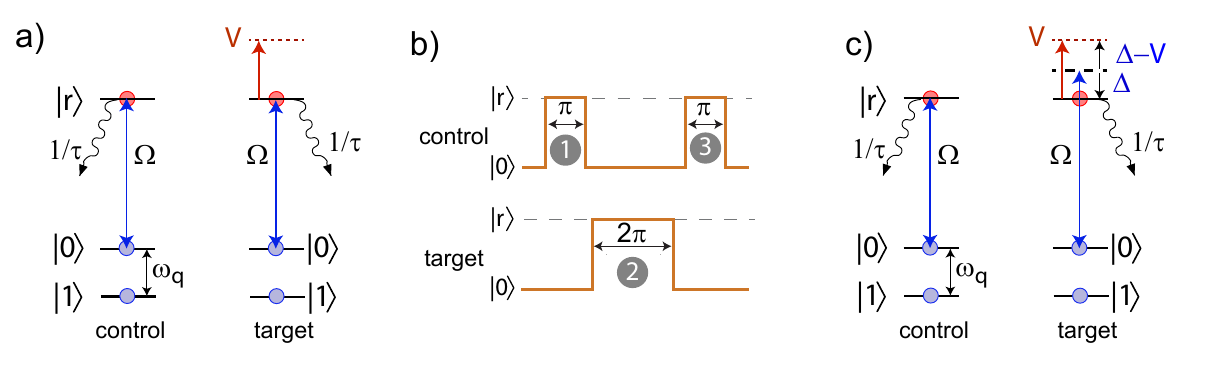}
 \vspace{-.2cm}
  \caption{Protocol for a Rydberg $\sf CZ$ gate. a) Atomic levels with qubits encoded in $\ket{0}, \ket{1}$ that are separated by $\omega_q$. State $\ket{0}$ is coupled to $\ket{r}$ with Rabi frequency $\Omega$. The Rydberg levels decay at rate $1/\tau$ and the two-atom interaction strength is $V$. b) A three pulse gate protocol is: 1) $\pi$ pulse on control qubit, 2) $2\pi$ pulse on target, 3) $\pi$ pulse on control. c) A modified gate protocol is the same as in a) and b) except that the $2\pi$ pulse on the target is detuned by $\Delta$.  }
\label{fig.gate}
\end{figure*}

We proceed in Sec. \ref{sec.pigate} by briefly reviewing the original $\pi-2\pi-\pi$ protocol shown in Fig. \ref{fig.gate} and the associated gate fidelity analysis. In Sec. \ref{sec.dangate} we  describe the new protocol, provide a simplified error analysis, and generalize the gate for arbitrary controlled phases. We generalize the protocol to different interaction strengths using optimal control in Sec. \ref{sec.optimalcontrol}. An enhanced asymmetric protocol that uses phase-modulated pulses on the target qubit to achieve robustness against variations in Rabi frequency and interaction energy is analyzed in Sec. \ref{sec.robustcontrol}.
We conclude with a comparison of the new protocol relative to fundamental limits in Sec. \ref{sec.discussion}.

\section{$\pi-2\pi-\pi$ Rydberg gate}
\label{sec.pigate}

The original Rydberg $\sf CZ$ gate protocol \cite{Jaksch2000} is based on a three pulse sequence. A $\pi$ pulse is applied to the control qubit, a $2\pi$ pulse is applied to the target qubit, and a final $\pi$ pulse is applied to the control qubit. Each pulse is resonant with the $\ket{0}\leftrightarrow\ket{r}$ transition\footnote{We adopt this convention to simplify phase calculations.} where the qubit basis is $\ket{0}, \ket{1}$ and $\ket{r}$ is the Rydberg state. In the idealized limit of $V, \omega_q\gg \Omega\gg 1/\tau$ where $V$ is the two-atom interaction strength, $\omega_q$ is the energy separation of $\ket{0}$ and $\ket{1}$,  $\Omega$ is the ground-Rydberg Rabi frequency, and $\tau$ is the Rydberg lifetime we obtain the mapping for control, target basis states $\ket{ct}$:
\begin{subequations}
\begin{eqnarray}
&&\ket{00}\rightarrow -e^{-\imath \phi_{00}} \ket{00},\\
&&\ket{01}\rightarrow -e^{-\imath \phi_{01}}\ket{01},\\
&&\ket{10}\rightarrow -e^{-\imath \phi_{10}}\ket{10},\\
&&\ket{11}\rightarrow e^{-\imath \phi_{11}}\ket{11}.
\end{eqnarray}
\end{subequations}
Here $\phi_{00},\phi_{01},\phi_{10},\phi_{11}$ are dynamical phases from differential Stark shifts on the qubit states due to the Rydberg pulses\rsub{. The particular values of the dynamical phases depend sensitively on the Rydberg laser excitation condition and are unimportant for the discussion here, as they only set required single-qubit phase corrections\footnote{Methods for measurement and correction of these single-qubit phase shifts are well-established in the literature, e.g Ref. \cite{Levine2019}.} and have no impact on entangling phase applied by the gate. T}he minus signs on the first three states are geometrical phases from the trajectory with area $2\pi$ in the $\{\ket{0},\ket{r}\}$ Hilbert space. When all three pulses use the same \rsub{excitation conditions} the dynamical phases can be expressed as 
$\phi_{00}=2\phi_{0}$, $\phi_{01}=\phi_{10}=\phi_0+\phi_1$, $\phi_{11}=2\phi_1$, with $\phi_0$, $\phi_1$ the single atom Stark shifts on states $\ket{0},\ket{1}$ during the gate. Applying a global phase rotation ${\sf R}_z(\phi_1-\phi_0)\otimes {\sf R}_z(\phi_1-\phi_0)$ and suppressing an irrelevant  global phase we obtain the canonical $\sf CZ$ gate: 
\begin{subequations}\label{eq:canonical}\begin{eqnarray}
&&\ket{00}\rightarrow  \ket{00},\\
&&\ket{01}\rightarrow  \ket{01},\\
&&\ket{10}\rightarrow  \ket{10},\\
&&\ket{11}\rightarrow - \ket{11}.
\end{eqnarray}\end{subequations}

In practice the ratios $V/\Omega$ and $\Omega/(1/\tau)$ are chosen to be large but finite, while $\omega_q\gg V$. This leads to the dominant error terms being those due to decay from the Rydberg state and incomplete blocking of the target atom rotation when the control atom is Rydberg excited. The errors are minimized by choosing the Rabi frequency to be 
$\Omega_{\rm opt}=(7\pi)^{1/3}\left(V^2/\tau \right)^{1/3}$ which leads  
 to a minimum gate error of \cite{Saffman2010}:
\begin{equation}
    \epsilon_{\pi-2\pi-\pi}=\frac{3(7\pi)^{2/3}}{8}\frac{1}{(V\tau)^{2/3}}.
    \label{eq.errorp2pp}
\end{equation}
Note that the minimum error scales as $\epsilon_{\pi-2\pi-\pi}\sim (V\tau)^{-2/3}$ whereas the best possible scaling for an arbitrary gate protocol is \rsub{shown by Doultsinos, Delakouras, and Petrosyan to be} $\epsilon_{\rm DDP}=(1+\pi/2)/ (V\tau)\simeq 2.57/(V\tau)$  \cite{Wesenberg2007,Doultsinos2025b}\rsub{; we emphasize that the scaling difference is not just a prefactor, but in the exponent of the gate duration $\tau$} . 

The highest fidelity that has been demonstrated with the $\pi-2\pi-\pi$ gate is ${\mathcal F}=1-\epsilon=0.89$ \cite{Graham2019} .
Detailed numerical analysis taking into account the finite spacing of Rydberg levels in Rb and Cs atoms \cite{XZhang2012} leads to an error floor in the absence of technical imperfections that is at best $\epsilon\simeq0.002.$ This can be improved on by using shaped pulses to suppress excitation of neighboring Rydberg states in which case errors as low as $\epsilon< 10^{-4}$ are theoretically achievable \cite{Theis2016b}. \rsub{However, the shaped pulse solutions only reach such high fidelity with Rydberg pulses an order of magnitude faster than those presented here, and several orders of magnitude stronger interactions.}

\section{Asymmetric $\pi-2\pi-\pi$ Rydberg gate}
\label{sec.dangate}

The error scaling of the gate can be improved by detuning the pulse on the target atom by an amount $\Delta=\omega-\omega_{r0}$ where $\omega$ is the laser frequency and $\omega_{r0}$ is the atomic ground-Rydberg frequency.  With this detuning the $\ket{10}$ state will return to the ground state provided 
\begin{equation}
t\sqrt{|\Omega|^2+\Delta^2}=2\pi
\label{eq.10}
\end{equation}
and the $\ket{00}$ state will return to the ground state provided 
\begin{equation}
t\sqrt{|\Omega|^2+(\Delta-V)^2}=2\pi.
\label{eq.00}\end{equation}
Here we assume $\Delta$ and $V$ are the same sign. 
In both cases we can achieve full return of the target atom to the ground state (neglecting Rydberg scattering) provided $\Delta=V/2$  and $t=2\pi/\sqrt{|\Omega|^2+V^2/4}.$ The amplitude factor imparted by the $2\pi$ pulse on the target atom when the control atom is not (is) excited is $e^{-i\phi_{(V)}}$ with
\begin{subequations}
\begin{eqnarray}
    \phi&=&\pi+ \frac{\Delta t}{2}= \pi+\frac{ \pi\Delta}{\sqrt{|\Omega|^2+\Delta^2}},\\
     \phi_V&=&\pi+ \frac{(\Delta-V) t}{2}=\pi +\frac{\pi( \Delta-V)}{\sqrt{|\Omega|^2+(\Delta-V)^2}}.
\end{eqnarray}
\label{eq.gphase}
\end{subequations}

Choosing $\Omega=\sqrt3 V/2$ and $t=2\pi/V$, we find $ \phi=3\pi/2,$ $ \phi_{V}=\pi/2$. The gate mapping on $\ket{ct}$ is
\begin{subequations}
\begin{eqnarray}
&&\ket{00}\rightarrow -e^{-\imath (\pi/2+\phi_{0c}+\phi_{0t})} \ket{00},\\
&&\ket{01}\rightarrow -e^{-\imath (\phi_{0c}+\phi_{1t})}\ket{01},\\
&&\ket{10}\rightarrow -e^{-\imath (\pi/2+\phi_{1c}+\phi_{0t})}\ket{10},\\
&&\ket{11}\rightarrow e^{-\imath (\phi_{1c}+\phi_{1t})}\ket{11}
\end{eqnarray}
\end{subequations}
where $\phi_{0 c/t}$ is the dynamical phase on state $\ket{0}$
for the control/target atoms and similarly for state $\ket{1}$. These phases are no longer identical for the control and target since the Rydberg pulse on the target atom is detuned. 
Applying single qubit rotations ${\sf R}_z(\phi_{1c}-\phi_{0c})\otimes {\sf R}_z(-\pi/2+\phi_{1t}-\phi_{0t})$ we obtain up to a global phase the canonical $\sf CZ$ gate of Eqs. (\ref{eq:canonical}). \rsub{We note that a related detuned pulse sequence which also achieves zero coherent rotation error was previously described in \cite{Shi2017}.}

Even though the interaction strength is finite, $V=2\Omega/\sqrt3$, there is no rotation error. To leading order the error is only due to Rydberg scattering. The integrated population in the Rydberg state, averaged over the computational basis states, is $P_{\rm r}=\pi(33+8\sqrt3)/(24V)$
which gives a scattering error of 
\begin{equation}
   \epsilon=\frac{P_{\rm r}}{\tau}=\left(\frac{11}{8}+\frac{1}{\sqrt3}\right)\frac{\pi}{V\tau}\simeq 2.39\times \epsilon_{\rm DDP}. 
\end{equation}
We see that the asymmetric gate protocol 
reaches within a factor of about three of the error limit. 

\begin{figure}[!t]
\centering
 \includegraphics[width=\columnwidth]{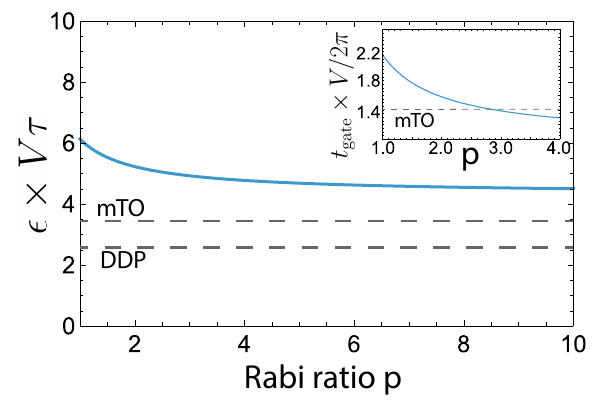}
 \vspace{-.2cm}
  \caption{Rydberg-scattering-limited gate error for the asymmetric protocol with $\Omega_c=p\Omega.$ The dashed line labeled DDP shows the bound of $\epsilon_{\rm DDP} V\tau=2.57 $ from \cite{Doultsinos2025b} and the line labeled mTO is the modified time optimal gate of Ref. \cite{Poole2025a}, Table II, gate 11. The inset shows the gate duration for the asymmetric and the modified time-optimal gates. }
\label{fig.error}
\end{figure}

We can reach closer to the error limit  with a simple modification to the protocol by driving the target atom with Rabi rate $\Omega$ and the control atom with Rabi rate $\Omega_c=p \Omega$, allowing for $p>1$. Since the control atom is not subject to blockade $p$ may be arbitrarily large up to practical limits set by available laser power and coupling to neighboring Rydberg states \cite{Theis2016b}. We then find 
\begin{equation}
 \epsilon=\left( \frac{11}{8}+\frac{1}{\sqrt3 p} \right)\frac{\pi}{V\tau}\rightarrow \frac{11\pi}{8}\frac{1}{V\tau}~{\rm for~ }p\rightarrow\infty.   
 \label{eq.asymp}
\end{equation}
The gate duration is 
\begin{equation}
t_{\rm gate}=\left( 1  +\frac{2}{\sqrt3 p}\right)\frac{2\pi}{V}\rightarrow  \frac{2 \pi}{V} ~{\rm for~ }p\rightarrow\infty.     
\end{equation}
The error and gate duration are plotted in Fig. \ref{fig.error}. At $p=1$ the error is $\epsilon=2.39 \epsilon_{\rm DDP}$  and the limit of $p\rightarrow\infty$ gives $\epsilon_{\rm min}=1.68\epsilon_{\rm DDP}$ for the asymmetric gate.  For comparison we note that the adiabatic Rydberg dressing gate \cite{Mitra2023} achieves $\epsilon/\epsilon_{\rm DDP}=2.45$, the modified time-optimal gate  \cite{Poole2025a}
achieves $\epsilon/\epsilon_{\rm DDP}=1.33$, and  the interaction gate \cite{Jaksch2000} in the limit of rapid excitation of both atoms to the Rydberg state achieves  $\epsilon/\epsilon_{\rm DDP}=1.22$. Thus the fidelity of the asymmetric gate introduced here is comparable to previous designs.

\subsection{Generalizations}
The asymmetric protocol may also be used to implement a general phase gate with the mapping $U={\rm diag}[1,1,1,e^{\imath\theta}]$ by adjusting the Rabi frequency $\Omega$ as $\Omega = V\cdot \sqrt{(\pi/\theta)^2-1/4}$. The target qubit pulse duration is adjusted accordingly to $t=2\theta/V$. Extending the target qubit pulse duration to $2n\theta/V$ produces a controlled $n\theta$ gate in which the state of the target qubit traverses $n$ loops on the $\ket{0}-\ket{r}$ Bloch sphere.

\begin{figure}[t]
 \includegraphics[width=\columnwidth]{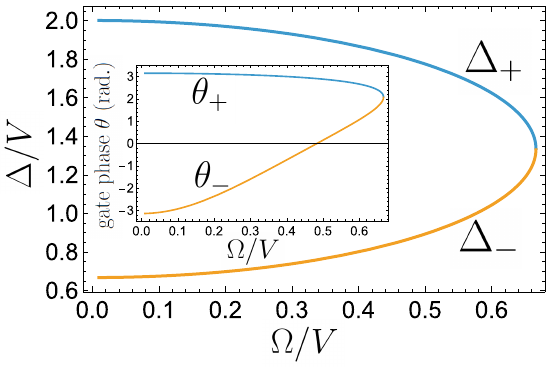}
 \vspace{-.7cm}
  \caption{General phase gate dependence on Rabi rate $\Omega$ from solutions of Eq. (\ref{eq.Deltapm}) for $n_1=2, n_2=1.$ The inset shows the corresponding gate phase.  }
\label{fig.general}
\end{figure}

Controlled phase gates can also be constructed for values of the detuning other than $\Delta=V/2$. In this case the detuning, Rabi frequency, and duration must be chosen so that $n_0$ ($n_V$), the number of Bloch-sphere loops traversed by the target qubit state in the absence (presence) of the Rydberg interaction, is an integer. We satisfy Eqs. (\ref{eq.10},\ref{eq.00}) at multiple $2\pi$ rotations as 
\begin{eqnarray}
t\sqrt{|\Omega|^2+\Delta^2}&=&2\pi n_0\\
t\sqrt{|\Omega|^2+(\Delta-V)^2}&=&2\pi n_V.
\end{eqnarray}
with $n_0,n_V$ integers. This leads to solutions for $n_0\ne n_V$:
\begin{equation}
\Delta_\pm=\frac{n_0^2V\pm\left[n_0^2n_V^2V^2-
(n_0^2-n_V^2)^2\Omega^2 \right]^{1/2} }{n_0^2-n_V^2}.
\label{eq.Deltapm}
\end{equation}
The target atom phases corresponding to Eqs. (\ref{eq.gphase}) are then given by 
\begin{eqnarray}
    \phi_\pm&=&n_0 \pi+\frac{n_0 \pi\Delta_\pm}{\sqrt{|\Omega|^2+\Delta_\pm^2}},\\
     \phi_{V,\pm}&=&n_V\pi +\frac{n_0\pi( \Delta_\pm-V)}{\sqrt{|\Omega|^2+\Delta_\pm^2}},
\end{eqnarray}
and the gate phase is 
\begin{equation}
    \theta_\pm=\phi_{V,\pm}-\phi_{\pm}.
\end{equation}
An example showing that $\theta$ can be tuned over the full range from $-\pi$ to $\pi$ is given in Fig. \ref{fig.general}.

\section{Time-optimal pulse designs}\label{sec.optimalcontrol}

\rsub{We now explore application of quantum optimal control to the gate discussed in the preceding sections.} By allowing a time-varying Rydberg laser phase during the target-qubit pulse \rsub{while otherwise leaving the pulse sequence unchanged}, a general family of gates is obtained. \rsub{This maintains the $\pi-2\pi-\pi$ structure of the gate while allowing generalization away from the ratio $\Omega/V=\sqrt{3}/2$.} \rsub{In this section we summarize our optimization methodology and then describe results for time-optimal control of the target-qubit pulse as $\Omega/V$ is varied, connecting results to limiting cases known from the literature. In the next section we explore gates in this family to identify waveforms that are robust to slow (e.g. gate-to-gate) variations in control parameters.} 

\rsub{Our explorations use} Gradient Ascent for Pulse Engineering (GRAPE)~\cite{Khaneja2005} to optimize a piecewise-constant Rydberg laser phase $\vec{\xi}$ during the target atom pulse, similar to optimizations for other gates~\cite{Jandura2022,Pagano2022}. \rsub{The vector $\vec{\xi}$ is defined for a target-qubit pulse of duration $\tau$ by $N=50$ values $\xi_n$, with $n=1, 2, ..., N$, describing the laser phase (relative to the co-rotating frame) for the interval between $t=(n-1) \tau/N$ and $n\tau/N$. GRAPE yields the fidelity gradient $\nabla_{\vec{\xi}} \mathcal{F}$ for the full gate sequence, which is used to maximize the fidelity and obtain the optimal target-qubit phase waveform $\vec{\xi}_\mathrm{optimal}$. For all the studies here we find that our optimizations produce phase trajectories corresponding to maximum instantaneous effective detunings arising from the phase modulation up to about $3\Omega$. Simulations are conducted through standard numerical evolution of the Schr\"odinger equation with Hamiltonian $H_T=H_{10}+H_{00} + \mathrm{c.c.}$, accounting for both control-qubit input states, with:}

\begin{align}
    \rsub{H_{10}} &\rsub{= \frac{\Omega}{2} e^{-i\xi}\ket{10}\bra{1r} -\frac{\Delta}{2}\ket{1r}\bra{1r} ,}\\
     \rsub{H_{00} }&\rsub{= \frac{\Omega}{2} e^{-i\xi}\ket{r0}\bra{rr} +\frac{1}{2}(V-\Delta)\ket{rr}\bra{rr}.}
\end{align}
\rsub{
To include decay out of the Rydberg state at rate $\Gamma$, we add an additional non-Hermitian term:
\begin{equation}
    H_{\mathrm{decay}}=-\frac{i\Gamma}{2}(\ket{r}\bra{r} \otimes I + I \otimes \ket{r}\bra{r}).
\end{equation}\label{eqn.rydbergdecay}}

\begin{figure}
\includegraphics[width=0.45\textwidth]{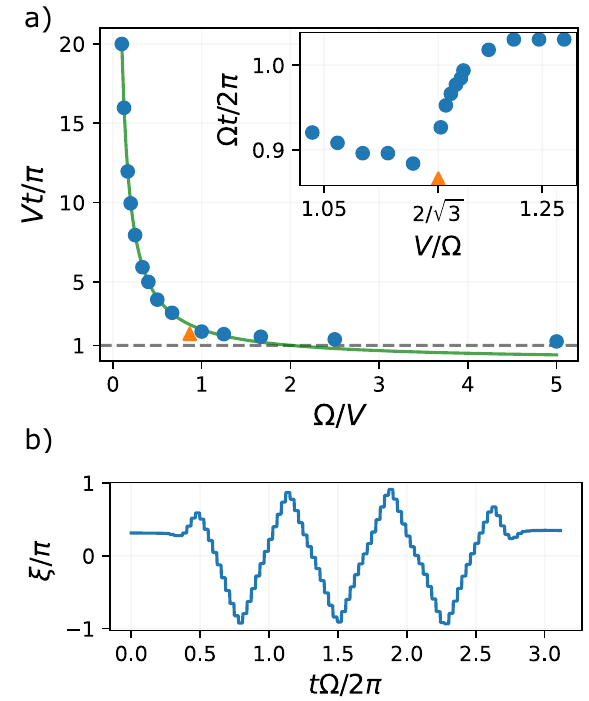}
\caption{Time-optimal solutions for the target pulse as the ratio $\Omega/V$ is varied. (a) GRAPE-optimized target pulse duration as a function of Rabi frequency $\Omega/V$ for a fixed interaction strength $V$. The limiting durations $\tau=2\pi/\Omega$ and $\tau=\pi/V$ discussed in the main text are shown by solid green and dashed grey lines, respectively. The duration of the analytic gate is shown by the orange triangle. Inset: Gate duration for fixed $\Omega$ as $V$ is varied. In this case the  analytic gate is the global minimum. (b)~Optimized gate waveform for $\Omega/V=5$, for which dynamics approximate the limit $\Omega\rightarrow\infty$. This waveform approximates the interaction gate and achieves effective free evolution with large detunings; here $|\Delta|_{\rm max}\approx 2.5\Omega$.
} \label{fig.optimalcontrol}
\end{figure}

\rsub{To use this formalism to search for time-optimal solutions, w}e use a binary search algorithm to find the shortest duration at which the error is below a target value (in the absence of decay mechanisms), which we set to be \rsub{$\eta=$} $10^{-8}$ \cite{buchemmavari2025}. \rsub{At each candidate time $\tau$, the target-pulse phase vector $\vec{\xi}$ is optimized by GRAPE.}
We do not include Rydberg decay in the dynamics for gate optimization as it does not make an appreciable difference in the optimal gate durations and waveforms~\cite{Jandura2022,Buchemmavari2024}.
Results of this optimization \rsub{as $\Omega/V$ is varied} are shown in Fig. \ref{fig.optimalcontrol}. For simplicity we describe gate durations in the limit $p\rightarrow\infty$, with $p=\Omega_c/\Omega$ as defined above. \rsub{Because optimal control is used only to optimize the pulse on the target qubit, the time-optimal durations depend on $p$ but the solutions do not.} 

We consider \rsub{two} notable limiting cases, \rsub{$\Omega/V\rightarrow\infty$ and $V/\Omega\rightarrow\infty$}. \rsub{For each, we track the smallest value of $\tau$ (found by binary search as described above) for which the GRAPE-optimal trajectory produces error below $\eta$ as the limit is taken}. First, if $V$ is fixed and \rsub{$\Omega \rightarrow \infty$ (while remaining fixed over the course of the gate)}, the gate duration approaches $\tau=\pi/V$ as $\Omega\rightarrow\infty$. In this limit the gate dynamics resemble the \rsub{``}interaction gate\rsub{'' (``Model A'', Ref. \cite{Jaksch2000})}, in which the target qubit in \rsub{both initial states $\ket{ct}= \ket{10}$ and }$\ket{00}$ (which is excited to $\ket{r0}$ by the first control-qubit pulse) is rapidly excited to the Rydberg state, the state is allowed to evolve for a time $\tau$, then the target qubit is rapidly de-excited. The free evolution can be achieved by setting $\Omega=0$; in our \rsub{optimization $\Omega$ is fixed over the course of the gate, but the optimization converges to a trajectory in which ground-Rydberg coupling is suppressed and phase evolution is canceled through switching of the detuning between large positive and negative values}, as shown in Fig. \ref{fig.optimalcontrol}b.

\begin{table}[!t]
\centering
\begin{tabular}{| c| c| c| }
\hline
   Fixed parameter & $\Omega$ & $V$ \\ \hline
   Analytic gate is optimal & Globally & Locally
 \\ \hline
 Limit & $V\rightarrow\infty$ & $\Omega\rightarrow\infty$ \\  \hline
 Asymptotic duration & $2\pi/\Omega$ & $\pi/V$  \\ \hline
 Asymptotic classification &$\pi-2\pi-\pi$ gate & Interaction gate \\ \hline
 Duration of analytic gate & $\sqrt{3}\pi/\Omega$ & $2\pi/V$ \\ \hline
\end{tabular}
\caption{A comparison of the duration of the analytic, constant-Rydberg-phase gate to asymptotic gates. We take $p\rightarrow\infty$, which has no impact on target-qubit dynamics and only affects total gate duration. The analytic gate is time-optimal when $\Omega$ is fixed and $V$ may be varied. It is locally optimal for fixed $V$ and variable $\Omega$, but longer than the interaction gate in the limit $\Omega\rightarrow\infty$. 
}
\label{tab.gatecomparison}
\end{table}

Second, if $\Omega$ is fixed and \rsub{$V\rightarrow\infty$}, then the gate duration asymptotically approaches $2\pi/\Omega$ as $V\rightarrow\infty$; this is the original $\pi-2\pi-\pi$ gate~\cite{Jaksch2000}. Table \ref{tab.gatecomparison} presents a comparison of these limiting cases with the analytic gate discussed above. 
We find that, within the family of gates considered here, an interaction strength of $V=2\Omega/\sqrt{3}$ optimally makes use of fixed available Rabi frequency for a duration of $\tau=5.441/\Omega$ (see Fig.~\ref{fig.optimalcontrol}a inset). This point also yields the time-optimal atomic separation for a fixed Rabi frequency. Similar speedups at moderate interaction strength have been observed in other Rydberg gates~\cite{Buchemmavari2024,Giudici2025}. An important caveat is that taking the limit $p\rightarrow\infty$ invalidates direct comparison to the time-optimal symmetrically addressed gate, which achieves total duration of $\tau=7.612/\Omega$ in the limit $V\rightarrow\infty$.

\section{Robust pulse designs}
\label{sec.robustcontrol}

In addition to optimizing gate speed for a fixed set of parameters, quantum control techniques can be employed to design robust gates. Here we consider gates that maintain high fidelity despite static but unknown errors in specific parameters. 
We use robust control to design gates that are resilient to variations in Rabi frequency $\Omega$ and interaction strength $V$. The optimization is performed at $\Omega = \sqrt{3} V/2$ and we assume $p\rightarrow\infty$ to simplify the numerics. 

Modifying the phase waveform of the pulse on the target qubit to obtain a robust gate inevitably requires an increase in the duration of the pulse relative to the time-optimal case, and the integrated Rydberg population increases accordingly. Therefore, finding concrete solutions requires assigning values to the Rydberg lifetime and to the scale of the variation to which robustness is desired. In our studies we take as a representative example $\Omega/\Gamma = 2\pi \times 150$, where $\Gamma$ is the decay rate out of the Rydberg state; this ratio roughly corresponds e.g. to $\Omega/2\pi=1$ MHz and $\Gamma=1/143\,\mu$s for Rydberg principal quantum number $n=70$ in $^{133}$Cs at $T=300$ K~\cite{Beterov2009}. Robustness against variation of a noisy Hamiltonian parameter with nominal value $q$ is obtained by optimizing the phase-modulation vector $\vec{\xi}$ to maximize the quantity $2{\mathcal F}(q) + {\mathcal F}(0.95q)+{\mathcal F}(1.05q)$ (other gate parameters are stable and are suppressed as arguments for simplicity), where we have chosen a 5 \% scale of variation for optimization~\cite{anderson2015prl,Mohan2023,Buchemmavari2024}.

\begin{figure}[!t]
\centering
\includegraphics[width=0.47\textwidth]{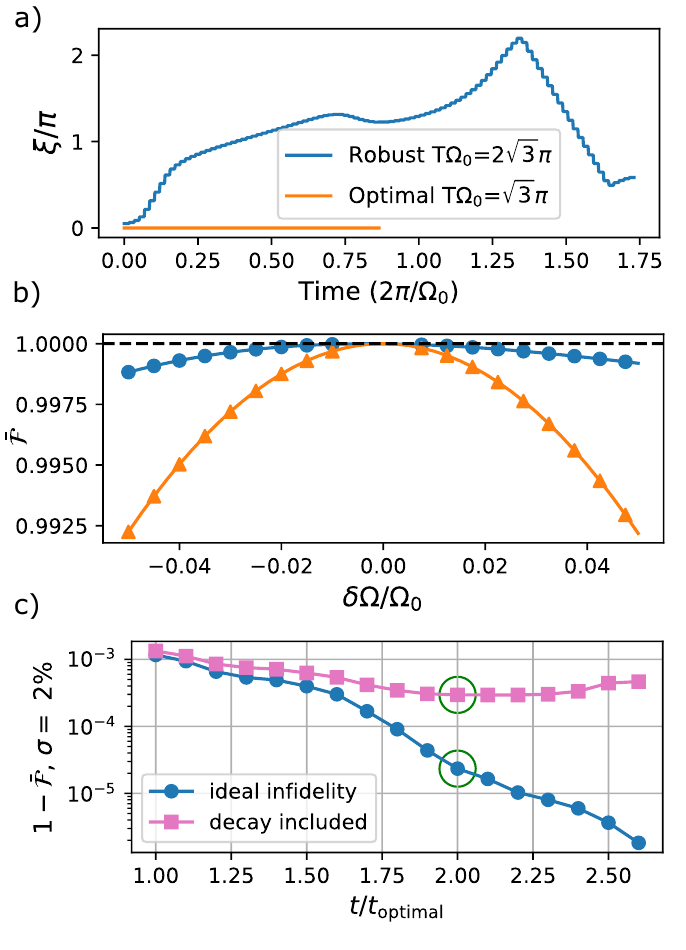}
\caption{
Properties of target atom phase waveforms for a CZ gate designed to be robust against Rabi frequency fluctuations. a) The phase waveform for the analytic (orange, flat), and robust (blue) pulses. The robust pulse is twice as long as the analytic pulse and exhibits significant phase modulation. The maximal detuning corresponds to $|\Delta|_{\rm max}\approx2\Omega_0$. 
b) Gate fidelity without Rydberg decay as a function of fractional Rabi frequency error ($\delta\Omega/\Omega_0$) for both pulses. The robust pulse maintains high fidelity across the full $\pm5\%$ range, with roughly sevenfold reduction in gate error at the extremes. 
c) Average gate error as a function of gate duration for Rabi-frequency-robust control, for normally distributed Rabi errors as described in the main text with $\sigma=2 \,\%$. As the gate duration increases, robust control achieves lower average gate error (Eq. \ref{eqn.averageinfidelity}) in the absence of decay (blue circles). The improvement is reduced when Rydberg decay is included (pink squares). At  $t/t_{\mathrm{optimal}}=2$ (shown in green circles), the gain from robustness is balanced by error from decay, resulting in a plateau followed by an eventual rise in gate error.}
\label{fig.Rabirobust}
\end{figure}

Fig. \ref{fig.Rabirobust} summarizes optimization for robustness against Rabi frequency variation. We find that the gate duration that provides the optimal compromise between minimization of fidelity variation over $\pm5\%$ and minimization of Rydberg-decay error is $t\approx2t_{\rm opt}$, with $t_{\rm opt}=2\pi/V$ for the analytic gate. As the ratio $\Omega/\Gamma$ increases, the optimal gate duration relative to $2\pi/\Omega$ increases and robustness improves~\cite{Tsai2025}. To further quantify robustness, we define an average weighted gate error over a Gaussian distribution of Rabi frequencies $\Omega$, centered at $\Omega_0$, with variance $\sigma^2$:

\begin{align}\label{eqn.averageinfidelity}
    1-\bar{\mathcal{F}} &= \int p(\Omega)(1-\mathcal{F}(\Omega))d\Omega \\
    p(\Omega) &= \frac{1}{\sqrt{2\pi\sigma^2}}\exp\left[{-\frac{1}{2}\left(\frac{\delta\Omega/\Omega_0}{\sigma}\right)^2}\right].
\end{align}
Figure \ref{fig.Rabirobust}c depicts the dependence of this error on the gate duration.

\begin{figure}
\centering
\includegraphics[width=0.48\textwidth]{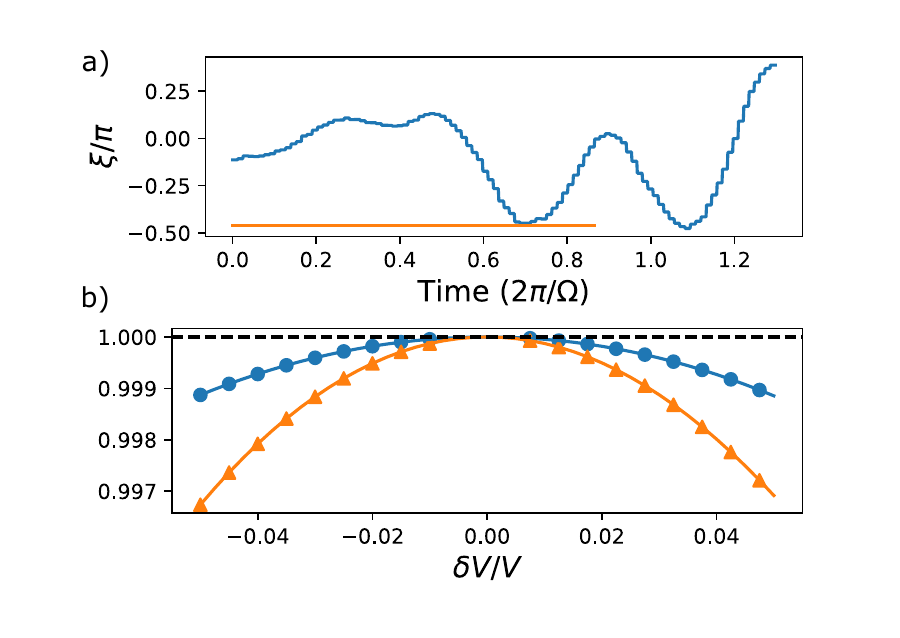}
\caption{
Properties of target atom phase waveforms for a CZ gate designed to be robust against interaction strength fluctuations. a) The phase waveform for the analytic (orange, flat), and robust (blue) pulses. The robust pulse is 50\% longer than the analytic pulse and exhibits significant phase modulation, with a maximum instantaneous detuning of $|\Delta|_{\rm max}\approx 2.8\times\Omega$.  b) Gate fidelity as a function of fractional interaction strength error ($\delta V/V$) for both pulses. The robust pulse maintains higher fidelity across the full $\pm5\%$ range, with roughly threefold reduction in gate error at the extremes. 
} \label{fig.vrobust}
\end{figure}

Fig. \ref{fig.vrobust} summarizes optimization for robustness against variation in the interaction strength $V$. Due to the dependence of the Rydberg interaction on inter-atomic distance as $V=r^{-\alpha}$ with $\alpha=3$ (dipole-dipole interaction) or $\alpha=6$ (van der Waals interaction), interaction can vary strongly for realistic noise in the inter-atomic separation. This is especially impactful in the regime where $\Omega\sim V$, as gate dynamics are much more sensitive to the particular ratio $\Omega/V$ than in the limit $V\rightarrow\infty$ where double excitation to $\ket{RR}$ is strongly suppressed over a wide range of values $V$. With $\delta V/V \approx -\alpha \delta r/r$, one finds for example that for atoms separated by 6 microns,  a plausible 50 nm spacing error (as may arise from finite temperatures in optical tweezers) produces $\delta V/V \approx 5 \%$. For optimization against this variation with $\Omega/\Gamma=2\pi\times150$, we find the optimal duration to be $t\approx 1.5 t_{\rm opt}$, at which the gate error is reduced threefold at 5 \% interaction-strength error.
\rrsub{The effects of Rydberg decay on the performance of the interaction strength robust waveform are similar to the previous case seen in Fig. \ref{fig.Rabirobust}c. There is a trade-off between using longer, and more robust waveforms and suffering increased Rydberg decay error from the increase in integrated Rydberg population during the gate. The details of the trade-off are highly dependent on the specific values of Rabi frequency and interaction strength.}

\section{Discussion}
\label{sec.discussion}

 A wide range of Rydberg gate protocols have been proposed and  implemented, although only a few of them have a gate error that scales as $1/(V\tau)$. The two-atom dark state gate \cite{Petrosyan2017} reaches 
 $\epsilon=14.8 \epsilon_{\rm DDP}$. The time-optimal gate reaches $\epsilon=11.7 \epsilon_{\rm DDP}$ for the parameters analyzed in Ref. \cite{Jandura2022}. It was shown in Ref. \cite{Poole2025a} that modifying the time-optimal profile by smoothly apodizing the sinusoidal phase variation could reach an error as low as $\epsilon=1.33 \epsilon_{\rm DDP}$. The protocol analyzed here reaches $\epsilon=2.39 \epsilon_{\rm DDP}$ for equal Rabi frequencies on control and target atoms and $\epsilon= 1.68\epsilon_{\rm DDP}$  when the control atom Rabi rate is much higher than that on the target atom.

 An additional feature of the asymmetric gate is that it is both fast and long range due to the ability to reach high fidelity with $\Omega \sim V$. In terms of the Rabi frequency the duration of the gate is $t_{\rm gate}\Omega=\sqrt{3}\pi + 2\pi/p.$ For $p=1$ this is about 50\% slower than the time-optimal gate, which requires $t_{\rm gate}\Omega \simeq 7.6$ \cite{Jandura2022}, but it is faster than the modified time-optimal gate for $p\ge2.9$ (see Fig. \ref{fig.error}). 
 
 The operating distance of the gate can be characterized by the error scaling relative to $\epsilon_{DDP}$, as a smaller scaling prefactor yields similar error at larger distance. The range of the asymmetric gate is larger than that of the time-optimal gate and similar to that of the modified time-optimal gate. \rsub{To give an explicit example, consider the Cs $82s_{1/2}$ state that was used in \cite{Maller2015} to implement the standard $\pi-2\pi-\pi$ gate at a distance of $7.6~\mu\rm m$. With the asymmetric protocol presented here the operation range of the gate can be extended to $10.2~\mu\rm m$ with a lifetime-limited error of $\epsilon=0.0012$. This follows from Eq. (\ref{eq.asymp}) with  $p=2$, $V=2\pi\times 4~\rm MHz$, and $\tau=203~\mu\rm s$. To reach the same lifetime limited error with the standard  $\pi-2\pi-\pi$ gate (see Eq. (\ref{eq.errorp2pp})) would require $V=2\pi\times 100~\rm MHz$, which implies an interaction distance of $r=6 ~\mu\rm m$. Extending the interaction distance from 6 to 10.2 $\mu\rm m$ implies an increase in interaction volume, and the number of qubits that can be connected,  of $2.9\times$ in 2D and $4.9\times$ in 3D.} 
 Relaxation of the  requirement of strong blockade facilitates logical two-qubit gates and quantum memory based on non-local low density parity check codes without atom motion \cite{Poole2025a,Pecorari2025}.

  On the other hand gates that do not operate with strong blockade are intrinsically less robust to variations in atom spacing and interaction strength, as well as being susceptible to force-induced heating due to double-excitation to $\ket{rr}$. As has been discussed elsewhere (see Eq. (45) in \cite{Robicheaux2021}) the gate error arising from forces between doubly-excited states scales as $T_a V^2 P_{\rm rr}^2 /(R^2 \omega^2). $ Here $T_a$ is the atomic temperature, $R$ is the atomic separation, $P_{rr}$ is the integrated $\ket{rr}$ population, and $\omega$ is the trap vibrational frequency along the interatomic coordinate. Although this 
  error can be reduced by increasing $\omega$, doing so increases the errors due to photon recoil. Full optimization of gate parameters is a constrained problem that is dependent on the actual physical implementation, and is outside the scope of this work.

Finally we note that the asymmetric protocol presented here was demonstrated with Cs atoms as reported in \cite{Cole2023}. The $66s_{1/2}$ level was excited using a two-photon approach with atoms spaced by $R\sim6~\mu\rm m$ giving an interaction strength of $V=2\pi\times 6~\rm MHz$. A gate fidelity as high as 0.964 was achieved, likely limited by technical errors due to laser noise and finite atom temperature.

\section{Acknowledgments}

VB thanks Ivan H. Deutsch for fruitful discussions and support throughout this project. VB also acknowledges support from the Quantum New Mexico Institute (QNM-I).

\bibliography{qc_refs,optics,saffman_refs,rydberg,atomic,thispaper}

\end{document}